\documentclass[fleqn,usenatbib]{mnras}

\usepackage{newtxtext,newtxmath}
\usepackage[T1]{fontenc}

\DeclareRobustCommand{\VAN}[3]{#2}
\let\VANthebibliography\thebibliography
\def\thebibliography{\DeclareRobustCommand{\VAN}[3]{##3}\VANthebibliography}

\usepackage{graphicx}	
\usepackage{amsmath}	
\usepackage{xspace}
\usepackage{tikz}
\usepackage[normalem]{ulem}
\usepackage[referable]{threeparttablex}

\newcommand{\pcube}{{\sc{pcube}}\xspace}
\newcommand{\xspec}{{\sc xspec}\xspace}

\newcommand{\ixpeobssim}{\mbox{{\sc ixpeobssim}}\xspace}

\newcommand{\ixpe}{{\it IXPE}\xspace}
\newcommand{\bepposax}{{\it BeppoSAX}\xspace}
\newcommand{\integral}{{\mbox{\it INTEGRAL}}\xspace}
\newcommand{\polarlight}{{\it PolarLight}\xspace}
\newcommand{\swift}{{\it Swift}\xspace}
\newcommand{\nustar}{{\it NuSTAR}\xspace}
\newcommand{\nicer}{{\it NICER}\xspace}
\newcommand{\maxi}{{MAXI}\xspace}
\newcommand{\jmx}{{JEM-X}\xspace}
\newcommand{\jmxone}{{JEM-X1}\xspace}
\newcommand{\jmxtwo}{{JEM-X2}\xspace}
\newcommand{\isgri}{{ISGRI}\xspace}
\newcommand{\cygx}{\mbox{Cyg X-2}\xspace}
\newcommand{\scox}{\mbox{Sco X-1}\xspace}

\newcommand{\tbabs}{\texttt{tbabs}\xspace}
\newcommand{\tbfeo}{\texttt{tbfeo}\xspace}
\newcommand{\comptt}{\texttt{comptt}\xspace}
\newcommand{\diskbb}{\texttt{diskbb}\xspace}
\newcommand{\polconst}{\texttt{polconst}\xspace}
\newcommand{\gauss}{\texttt{gaussian}\xspace}
\newcommand{\relconv}{\texttt{relconv}\xspace}
\newcommand{\reflionx}{\texttt{reflionx}\xspace}
\newcommand{\relxillns}{\texttt{relxillns}\xspace}

\newcommand{\rfxconv}{\texttt{rfxconv}\xspace}

\title[Polarization of \cygx]{Accretion geometry of the neutron star low mass X-ray binary \cygx\  from X-ray polarization measurements}

\author[R. Farinelli et al.]{
R. Farinelli,$^{1}$\thanks{E-mail: ruben.farinelli@inaf.it}
S. Fabiani,$^{2}$
J. Poutanen,$^{3,4}$ 
F. Ursini,$^{5}$ 
C. Ferrigno,$^{6}$ 
S. Bianchi,$^{5}$ 
M. Cocchi,$^{7}$ 
F. Capitanio,$^{2}$ 
\newauthor
A. De Rosa,$^2$
A. Gnarini,$^{5}$ 
F. Kislat,$^8$ 
G. Matt,$^{5}$ 
R. Mikusincova,$^{5}$ 
F. Muleri,$^{2}$
I. Agudo,$^{9}$
L. A. Antonelli,$^{10,11}$
\newauthor
M. Bachetti,$^{7}$
L. Baldini,$^{12,13}$
W. H. Baumgartner,$^{14}$
R. Bellazzini,$^{12}$
S. D. Bongiorno,$^{14}$
R. Bonino,$^{15,16}$
\newauthor
A. Brez,$^{12}$
N. Bucciantini,$^{17,18,19}$
S. Castellano,$^{12}$
E. Cavazzuti,$^{20}$
S. Ciprini,$^{21,11}$ 
E. Costa,$^{2}$
E. Del Monte,$^{2}$
\newauthor
L. Di Gesu,$^{20}$
N. Di Lalla,$^{22}$
A. Di Marco,$^{2}$
I. Donnarumma,$^{20}$
V. Doroshenko,$^{23}$
M. Dov\v{c}iak,$^{24}$
S. R. Ehlert,$^{14}$
\newauthor
T. Enoto,$^{25}$
Y. Evangelista,$^{2}$
R. Ferrazzoli,$^{2}$
 J. A. Garcia,$^{26}$
S. Gunji,$^{27}$
K. Hayashida,$^{28}$
J. Heyl,$^{29}$
\newauthor
W. Iwakiri,$^{30}$
S. G. Jorstad,$^{31,32}$
V. Karas,$^{24}$
T. Kitaguchi,$^{25}$
J. J. Kolodziejczak,$^{14}$
H. Krawczynski,$^{33}$
\newauthor
F. La Monaca,$^{2}$
L. Latronico,$^{15}$
I. Liodakis,$^{34}$
S. Maldera,$^{15}$
A. Manfreda,$^{12}$
F. Marin,$^{35}$
A. P. Marscher,$^{31}$
\newauthor
H. L. Marshall,$^{36}$
I. Mitsuishi,$^{37}$
T. Mizuno,$^{38}$
C.-Y. Ng,$^{39}$
S. L. O'Dell,$^{14}$
N. Omodei,$^{22}$
\newauthor
C. Oppedisano,$^{15}$
A. Papitto,$^{10}$  
G. G. Pavlov,$^{40}$
A. L. Peirson,$^{22}$
M. Perri,$^{11,10}$
M. Pesce-Rollins,$^{12}$
\newauthor
P. O. Petrucci,$^{41}$
M. Pilia,$^{7}$
A. Possenti,$^{7}$
S. Puccetti,$^{11}$
B. D. Ramsey,$^{14}$
J. Rankin,$^{2}$
A. Ratheesh,$^{2}$
\newauthor
R. W. Romani,$^{22}$
C. Sgr\`o,$^{12}$
P. Slane,$^{42}$
P. Soffitta,$^{2}$
G. Spandre,$^{12}$
T. Tamagawa,$^{25}$
F. Tavecchio,$^{43}$
\newauthor
R. Taverna,$^{44}$
Y. Tawara,$^{37}$
A. F. Tennant,$^{14}$
N. E. Thomas,$^{14}$
F. Tombesi,$^{45,21,46}$  
A. Trois,$^{7}$
\newauthor
S. S. Tsygankov,$^{3,4}$
R. Turolla,$^{44,47}$ 
J. Vink,$^{48}$
M. C. Weisskopf,$^{14}$
K. Wu,$^{47}$
F. Xie,$^{49, 2}$ 
S. Zane$^{47}$ 
\newauthor 
\textit{\small Affiliations are listed at the end of the paper} 
}

\date{Accepted XXX. Received YYY; in original form ZZZ}

\pubyear{2022}

\begin{document}
\label{firstpage}
\pagerange{\pageref{firstpage}--\pageref{lastpage}}
\maketitle

\begin{abstract}
We report spectro-polarimetric results of an observational campaign of the bright neutron star low-mass X-ray binary \cygx simultaneously observed by \ixpe, \nicer\ and \integral. 
Consistently with previous results, the broad-band spectrum is characterized by a lower-energy component, attributed to the accretion disc with $kT_{\rm in} \approx$ 1 keV, plus unsaturated  Comptonization in thermal plasma with temperature $kT_{\rm e} = 3$ keV and optical depth $\tau \approx 4$, assuming a slab geometry.
We measure the polarization degree in the 2--8 keV band $P=1.8 \pm 0.3$ per cent  and polarization angle $\phi = 140\degr \pm 4\degr$, consistent with the previous X-ray polarimetric measurements by \textit{OSO-8} as well as with the direction of the radio jet which was earlier observed from the source.
While polarization of the disc spectral component is poorly constrained with the \ixpe data, the Comptonized emission has a polarization degree $P =4.0 \pm 0.7$ per cent  and a polarization angle aligned with the radio jet.
Our results strongly favour a spreading layer at the neutron star surface as the main source of the polarization signal. 
However, we cannot exclude a significant contribution from reflection off the accretion disc, as indicated by the presence of the iron fluorescence line.
\end{abstract}


\section{Introduction}
\label{sect:introduction}

The physics of X-ray binaries systems (XRBs) hosting either a neutron star (NS) or a black hole (BH), has been a long-time matter of study by theoreticians and observers.
Both spectral and temporal features have been investigated across decades with different space observatories, but some questions remain still unanswered.
One of the most debated topics concerns the shape and location of the region  which is responsible for the strong Comptonization feature observed in the X-ray spectra of these sources and dominating the spectral emission at high energies \citep[for a review, see][]{done2007}. 
Often this component has been attributed to a so-called ``Compton corona'' \citep[e.g.,][]{HM93,ps96,fiorito2004, balucinska2010}.
Several models and configurations have been proposed across years, but the issue has not yet been solved unambiguously.

Is the corona located above the accretion disc or is it the region between the truncated cold accretion disc and the compact object (hot flow) \citep{PVZ18}? 
How the geometry of the emitting region  is affected by the presence of an event horizon or a solid surface ? 
How do its properties  change with the mass accretion rate? 
In this context, X-ray polarimetry has long been considered the key new window of X-ray astronomy, and 
with the launch of the {\it Imaging X-ray Polarimetry Explorer} \citep[\ixpe,][]{Weisskopf2022}, there are now great expectations on the possibility to disentangle among degenerate parameters  which can equally well describe spectral and temporal properties of XRBs.

Among the NS low-mass XRBs (LMXBs), \cygx\ has long been studied because of its high X-ray flux $F_{\rm X}\sim (1-3)\times 10^{-8}$\,erg\,cm$^{-2}$\,s$^{-1}$.
It is classified as a Z-source from the shape of the tracks in the colour–colour and hardness-intensity diagram on the time-scales ranging from hours to days \citep{hk89}.
From the spectral point of view, Z-sources are characterised by parameters which remain relatively stable during the motion along the Z pattern \citep{disalvo2000,lavagetto2004, disalvo2001, disalvo02, farinelli09,balucinska2010}.  
The most noticeable transient feature is instead a hard X-ray excess above 30 keV which occurs when the source is in the Horizonthal Branch (HB)  and appears to be correlated to the episodic radio emission indicating the presence of a jet \citep{damico2001, paizis2006, farinelli09}.
Joint observations of \cygx\ with \swift/XRT and the VLBI at 5 GHz have shown the presence of a sub-relativistic jet when the source was in the HB \citep{spencer2013}. 
In particular, the radio observations of 2013 February 22 and 23 revealed a single core at 0.59 mJy  the first day, while in the second day the core emission had weakened in favour of a  bright head in the south-east direction, 141\degr\ east of north. 
Using broad-band \bepposax\ data (0.1--200 keV), \cite{disalvo02} described the spectrum of \cygx\ in terms of a soft multi-colour disc component plus unsaturated thermal Comptonization in a plasma with low temperature ($kT_{\rm e} \sim 3$ keV) and moderately high Thomson optical depth ($\tau \sim 5-10$, depending on the geometry).
This spectral modelling falls in the so-called `eastern-model' scenario, where the soft component is attributed to the accretion disc, while Comptonization originates in a hot boundary or spreading layer (BL/SL) close to  the NS.
By the BL we mean part of the accretion disc where the gas decelerates \citep[e.g.,][]{Popham01}, while by the SL we mean the gas layer at the NS surface, which can extend to high latitudes \citep{inogamov1999,  suleimanov2006}.

The eastern-model framework has gained stronger support from the  analysis of a sample of atoll- and Z-sources using Fourier-frequency resolved spectroscopy of the archival {\it RXTE} data \citep{gilfanov2003, Revnivtsev06,Revnivtsev13}, and it was shown that the SL spectrum resembles the Fourier-frequency resolved spectrum at frequencies of quasi-periodic oscillations. 

Recently, \citet[][hereafter L22]{ludlam2022} performed a spectral analysis of  \nicer and \nustar\ data at different positions of the source in its Z-track.
Using, among others,  a reflection model for reprocessed radiation in the accretion disc, they inferred the inner disc radius close to the innermost stable circular orbit and  rather stable when the source moves in the hardness-intensity diagram.
The model also provided an orbital inclination $i=$60\degr--70\degr, consistent with the optical results \citep{orosz1999}.

Optical polarimetric observations of \cygx\ were performed in the $U$, $B$ and $V$ bands by \citet{miramond1995}, who found statistically significant linear polarization.
Comparing the position in the Stokes plane with that of nearby stars, the authors concluded that part of the observed polarization is intrinsic to the source,  with a polarization degree (PD)  of 0.29$\pm$0.07 per cent  and a polarization angle (PA) of $113\degr\pm 7\degr$.
This is $4\sigma$ off from the jet direction, which is not surprising because of the uncertainty in the measurements of the interstellar polarization.
Previous X-ray polarimetric observations of \cygx in 1975 with {\it OSO-8} gave a marginal detection of PD=$5.0\pm1.8$ per cent  at PA$=138\degr\pm10\degr$, while the later observations in 1976--1977 gave null  result     \citep{Weisskopf76,Long80}.
Interestingly, the PA coincides with the position angle of the radio jet. 

If polarization is produced in the accretion disc dominated by electron scattering, the linear PD drops from 11.7 per cent  for edge-on systems ($i=90\degr$) to zero for a face-on observer at $i=0\degr$  \citep{Chandrasekhar47,Sobolev49,chandrasekhar1960,Sobolev63} with the dominant direction of the electric field oscillations along the disc plane (i.e. perpendicular to the disc rotation axis).
Because at higher orbital inclination the object is expected to have a higher polarization in systems like LMXBs, \cygx, with a likely inclination angle of $i \sim 60\degr$, is a good of candidate to be observed by \ixpe. 
On the other hand, previous {\it OSO-8} results contradict this simple interpretation, because the polarization was aligned with the jet, which is likely perpendicular to the disc. 
Thus it is important to determine the X-ray PD and PA and their energy dependence with a higher precision, which is now possible with \ixpe. 

\section{Observations and data reduction}

\subsection{\ixpe}

Observations with \ixpe\  have been carried out from 2022 April 30 11:27 to 2022 May 02 11:05 UT for about 75.5 ks of net exposure time.
Data reduction and analysis were performed  by using the \ixpeobssim \footnote{\url{https://ixpeobssim.readthedocs.io/en/latest/?badge=latest}} software v26.3.2 \citep{Baldini} and \textsc{heasoft} tools v6.30.1.\footnote{\url{https://heasarc.gsfc.nasa.gov/docs/software/heasoft/}} 
The \ixpeobssim tools include \textsc{xppicorr} to locally apply the energy calibration with in-flight calibration sources (which at the time of writing was not included in the pipeline producing publicly available data), \textsc{xpselect} to filter data and \textsc{xpbin} to apply different binning algorithms for generating images and spectra. Rebinning and spectro-polarimetric analysis was performed by means of \textsc{heasoft} \textsc{ftools} including \xspec.\footnote{\xspec was used by means of the \textsc{pyxspec} interface. 
See the documentation at \url{https://heasarc.gsfc.nasa.gov/xanadu/xspec/python/html/index.html}} 
Source and background regions where selected from the image of each of the three detector units (DU), with the source centred in a circular region of 60\arcsec\ in radius. 
The background is extracted from an annular region with the inner and outer radii of 180\arcsec\ and 240\arcsec, respectively. 
The background counts are  negligible with respect to the source (about $\sim 0.2$ per cent).
The data analysis was performed following the unweighted method.\footnote{In the unweighted analysis method, equal weights are assigned to each photo-electron track, regardless of its shape.} 

The normalized Stokes parameters $Q/I$ and $U/I$, corresponding PD and PA, as well as their uncertainties are calculated using the \pcube\ binning algorithm of \ixpeobssim, which assumes that they are uncorrelated and that PD and PA are independent \citep{Kislat2015}.
We also compared the results of the polarimetric analysis obtained with both \xspec\ and \pcube. 
While \xspec\ requires the definition of a spectro-polarimetric model, \pcube\ is a model-independent tool that computes the polarization parameters only on the basis of the properties of the detected photons and instrument response matrices.

\begin{table} 
\caption{Log of the observations of \cygx.}
\label{tab:log_obs}
\begin{tabular}{lcc} 
\hline
Instrument/Satellite & $T_{\rm start}$ &  $T_{\rm stop}$\\
\hline
\hline
\nicer (Obs. 1)  & 2022--04--30 02:07:20 & 2022--04--30 23:58:40\\
\nicer (Obs. 2) & 2022--05--01 01:13:29 & 2022--05--01 23:08:00\\
\ixpe & 2022--04--30 11:27:00 & 2022--05--02 11:05:00\\
\integral & 2022--04--30 21:44:31 & 2022--05--01 12:03:00\\
\hline
\end{tabular}
\end{table}

The uncertainties from the \xspec  analysis are computed with the \texttt{error} command of \xspec for one parameter of interest.
It is worth noting that because the PD and PA are actually not independent, then a more appropriate way to report results is by means of contours representing, e.g., $68.27$,  $95.45$ and $99.73$ per cent confidence levels of the joint measurement of the two parameters.
The contour plot with \pcube\ are derived following \citet{Weisskopf2010}, \citet{Strohmayer2013}, and \citet{Muleri2022} by using the parameters obtained by the algorithm itself. 
In \xspec\ on the other hand, this can be achieved by using the \texttt{steppar} command by specifying the two parameters of interest (in this case, PD and PA).


\subsection{\integral and \nicer}
\label{sect: integral_nices}

We extract data of the public \integral  \citep{Winkler2003} target-of-opportunity observation carried out from 2022 April 30 21:44 to 2022 May 01 12:03 UT using the latest available calibration and software (Offline Scientific Analysis v11.2) from the multi-messenger online analysis platform.\footnote{\url{https://www.astro.unige.ch/mmoda/}} 
The \integral spacecraft has on board four co-aligned instruments: we used data from the \isgri\ coded-mask imager \citep{lebrun2003}, which is sensitive above $\sim$25\,keV in this late phase of the mission, and \jmx\ \citep{Lund2003}, consisting of two identical detectors sensitive in the 3--30\,keV band.
We first extract images for \isgri\ and \jmxone which are obtained as a mosaic from different dither pointing of the satellite.
This intermediate step is necessary to select the sources present in the field of view  of each instrument in addition to \cygx, and which contribute as `background' for the extraction of the spectra and light curves: for \isgri\ they are Cyg X-3, EXO 2030+375, and SS Cyg while for \jmx\ it is only SS Cyg,     because of its smaller field of view.
It is worth noting that \cygx\ is detected in the \isgri\ image with poor statistical significance, on the contrary to what happen in the \jmxone mosaic image.

We then extracted the \jmx spectra from the two units in sixteen standard energy bins and added a 5 per cent  systematic uncertainty to them. 
The \isgri\ spectrum is extracted at the natural resolution of 256 bins and then grouped in 20 equally spaced logarithmic bins from 25 to 200\,keV; an additional 1.5 per cent systematic error is added in quadrature.\footnote{\url{https://gitlab.astro.unige.ch/reproducible_INTEGRAL_analyses/cygx-2.git}.}

Two observations of the source were performed also by \nicer\ \citep{Gendreau16} on 2022 April 30 (ObsID 5034150102) and 2022 May 1 (ObsID 5034150103), with a total exposure of 8.1 ks. 
We reduced the \nicer~data using \textsc{heasoft} v6.30 and the \textsc{nicerl2} task and applying standard calibration and screenings, with \textsc{caldb} v20210707. 
The presence of sharp changes in the instrument effective area  due to gold edges on the mirror coating, which may be seen in residuals, is a known issue of the instrument, and would suggest to perform spectral analysis starting from about 2.5 keV \citep{miller2018}.
In order to achieve a trade-off between spectral coverage and instrumental uncertainties, we set the lower threshold for \nicer\ at 1.5 keV.

\section{Results}

\subsection{Temporal properties}
\label{sect:temporal properties}

\begin{figure}
\centering 
\includegraphics[width=1.1\columnwidth]{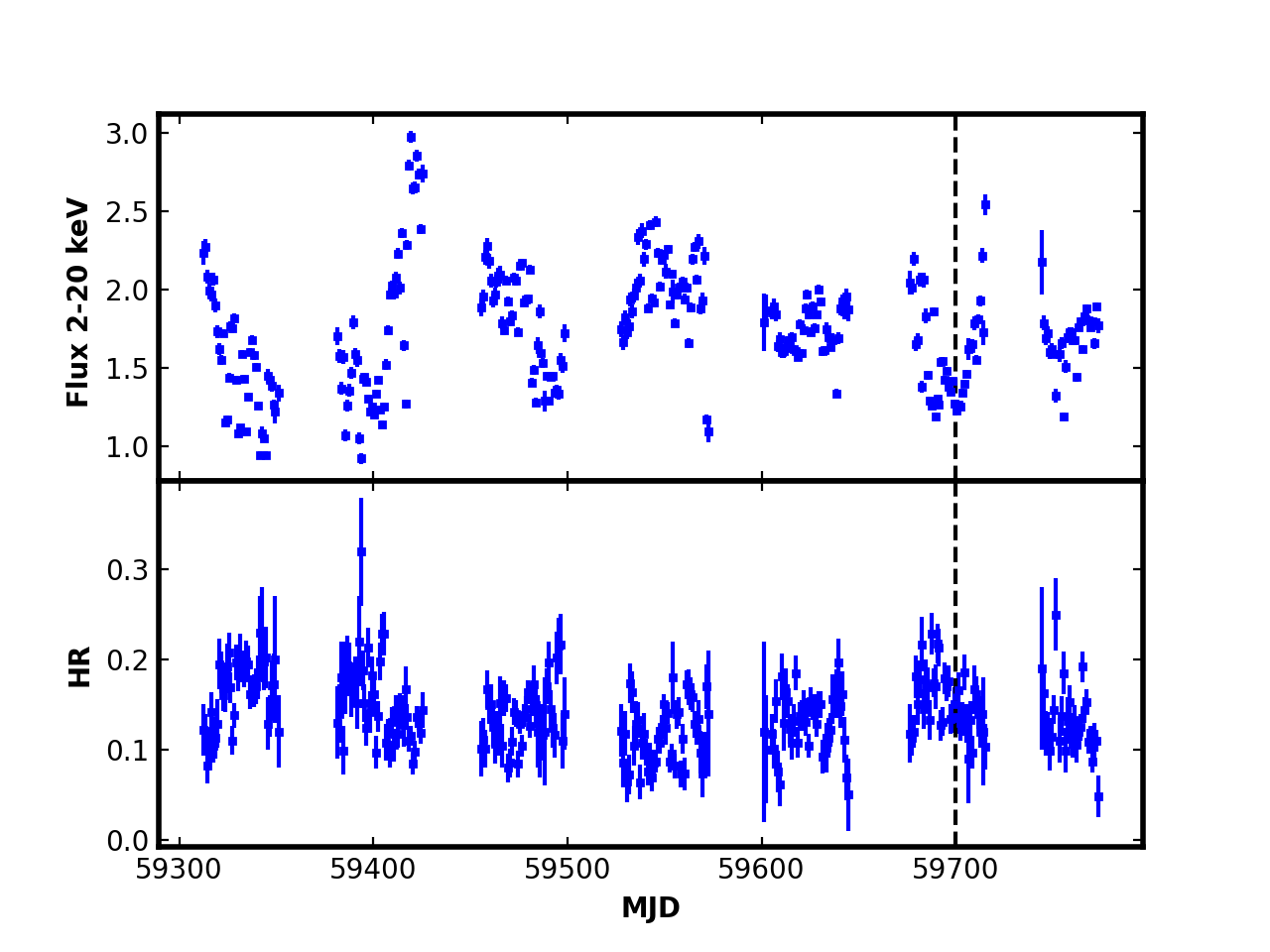} 
\caption{\textit{Top panel}: daily average \maxi\ light curve of \cygx\ in the energy band  2--20 keV 
in units of photons cm$^{-2}$~s$^{-1}$. \textit{Bottom panel}: hardness ratio between fluxes in the interval 10--20 keV and 4--10 keV, respectively.
The vertical dashed line corresponds approximately to the observations of \nicer, \ixpe\ and \integral 
(see Table~\ref{tab:log_obs}).}    
\label{maxi_lc}
\end{figure}

In Fig.~\ref{maxi_lc} we show the 2--20 keV photon flux and the hardness ratio (HR) of \cygx\ using data from the public \maxi\  \citep{Matsuoka09} archive.\footnote{\url{http://maxi.riken.jp/top/lc.html}} 
In particular, we consider the time starting one year before the observations of \nicer, \ixpe\ and \integral\ (MJD=59699--59700).
The source flux varied by about a factor of three over months and at the time of the observations  it was close to its minimum.
The HR variations instead were less pronounced, remaining within a factor two. 
This is consistent with the fact that spectral changes in Z-sources remain fairly moderate even when they cross the full Z-pattern \citep[e.g.,][]{disalvo2000, damico2001}.

\begin{figure}
\centering 
\includegraphics[angle=-90,width=1\columnwidth]{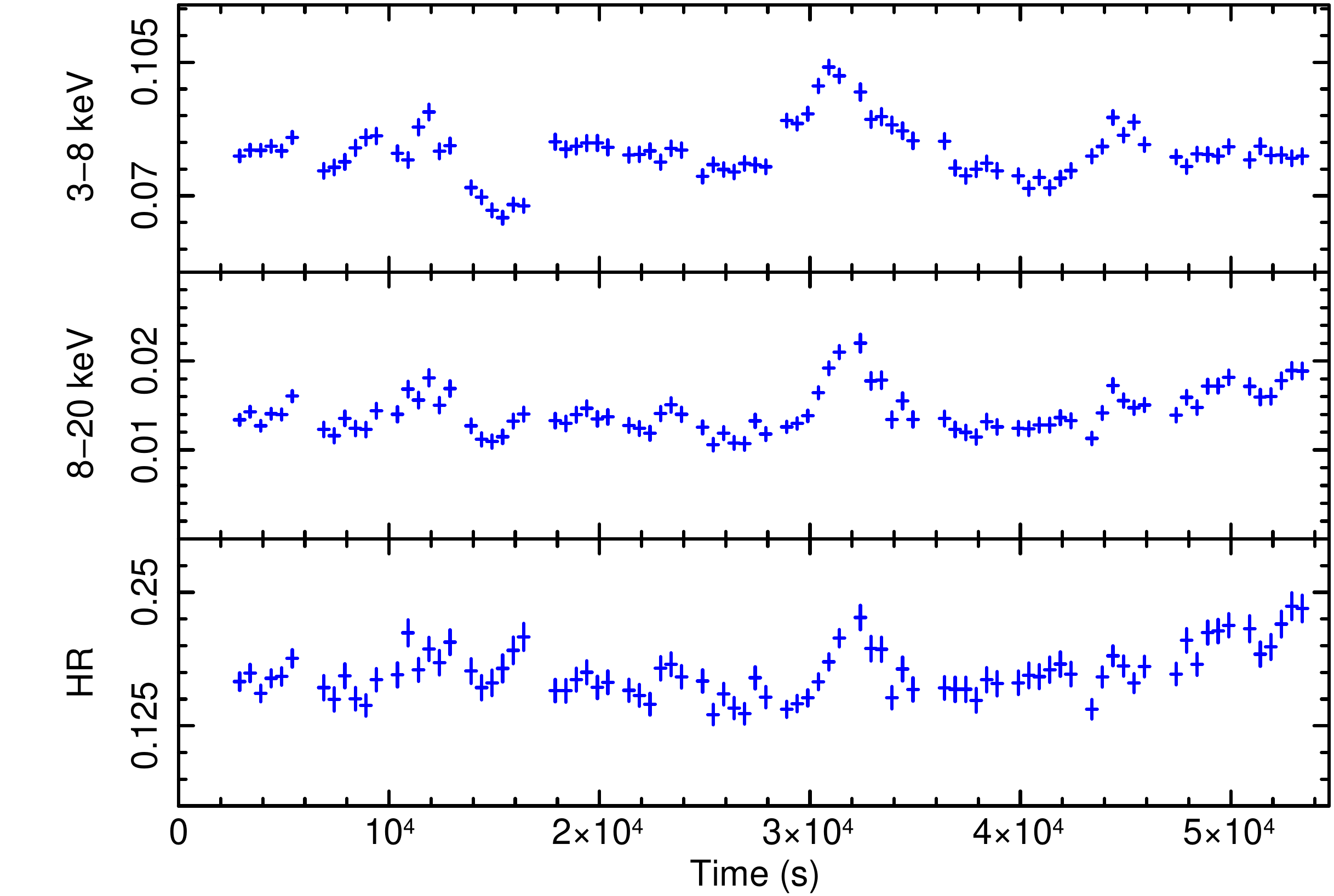}
\caption{Light curves of \cygx\ (in counts\ s$^{-1}$) in the soft and hard energy band of \jmxone, and their hardness ratio. 
Each bin corresponds to 500~s and time is measured from the beginning of the \integral\ observation.}
\label{jmx_lc}
\end{figure}

\begin{figure}
\centering 
\includegraphics[width=1\columnwidth]{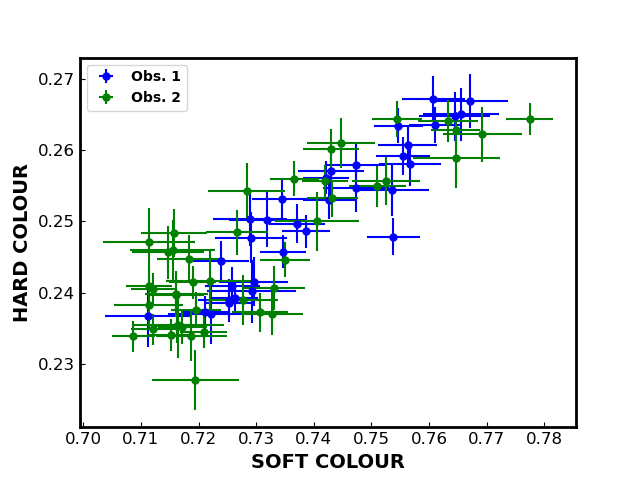}
\includegraphics[width=1\columnwidth]{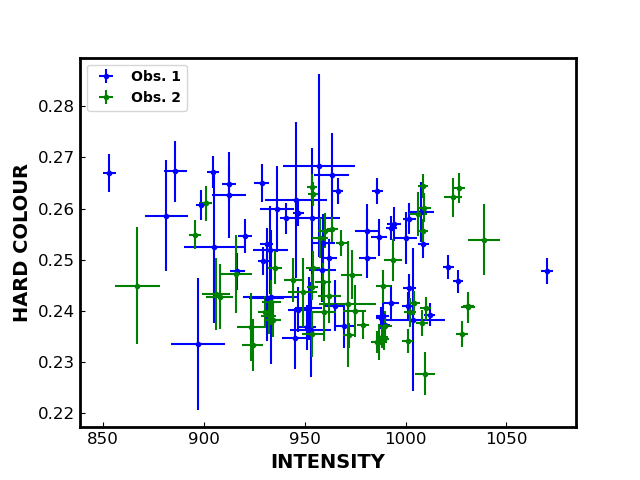}
\caption{\emph{Upper panel}: \nicer\ colour-colour diagram of both observations of \cygx. The hard colour and soft colours are defined as the count rate ratios 5--8 keV/3--5 keV and 3--5 keV/2--3 keV, respectively. \emph{Lower panel}: colour-intensity diagram with hard colour defined  in the same way, while intensity is the count rate in the 2--8 keV band. Each bin corresponds to a time interval of 150~s.}
\label{nicer_ccd}
\end{figure}

The time variability on a shorter time-scale was first investigated using \jmxone, which allows to cover the energy range up to about 20 keV. 
We selected two energy bands (3--8 keV and 8--20 keV) plotting the count rate behaviour as well as their  HR over time.
As shown in Fig.~\ref{jmx_lc}, both soft and hard band demonstrate a qualitatively similar pattern, albeit the peak in the hard band is relatively higher, being reflected in the HR.
Motivated by the overlapping between  the energy range of \jmx\ with that selected by L22 to produce hardness intensity diagrams (HIDs) of previous \nustar\ observations of \cygx, we also tried to make a corresponding HID, but because of the statistics we did not observe a clear evolution along the Z-like pattern -- just something which is reminiscent of an Upper Normal Branch. 
As a next step, we produced both a colour-colour diagram (CCD) and HID from the two \nicer\ observations (see Fig.~\ref{nicer_ccd}). 
The source pattern in the CCD reminds a Normal Branch (NB), albeit it cannot be excluded that its upper right portion does represent part of the Horizontal Branch (HB).
Similarly,  the CCD of the \cygx\ from observations performed in 1996 and 1997 with \bepposax\ appeared to be an extended NB, while the first data set revealed its HB nature when plotted on a HID \citep{farinelli09}.
Unlike the work  mentioned above, the intensity variations in the \nicer\ data were less pronounced, showing no definite pattern in the HID.  
Qualitatively similar CCD and HID have been obtained in the same energy bands with \ixpe, but with larger errors owing to a smaller effective area of the instrument.
Having neither simultaneous radio observations nor information about the presence of a high-energy X--ray tail, which are characteristic features of the HB, we shall cautiously consider the source as having traced the NB.

\subsection{Spectral analysis}
\label{sect:spectral analysis}

When considering all instruments which performed the pointing on \cygx\ (see Table~\ref{tab:log_obs}), the source has been monitored for a bit less than 2.5 d. 
This duration is nearly as long as the 1997 \bepposax\ observation, during which only the NB was traced out. 
The \integral\ observation ($\sim 50$ ks) has been performed nearly between the two \nicer pointings and fully within the \ixpe\ exposure, which lasted for about one day after the end of the \integral exposure.
If on one side, the time-resolved spectropolarimetry would be highly desirable to provide unique information about, e.g., the orbital phase dependence, from the other it is necessary to achieve a trade-off between source evolution and statistics.
In this context, having observed no dramatic source variations and, in particular, no flaring phase, we extracted a single average spectrum of the source for the whole observation from  \jmx, \isgri\ and \nicer.

We fitted the spectrum with a two-component model consisting of the multi-colour disc blackbody \citep[\diskbb,][]{mitsuda84} for the low-energy part and a Comptonization model \comptt\ \citep{titarchuk94} for higher energies.
For \comptt\ we select the flag which reports the value of the optical depth $\tau$ for the slab geometry.
The gap in the data between 20 and 30 keV together with the low energy resolution of \jmx, however, does not allow us to constrain simultaneously the electron temperature and optical depth, so we fixed $kT_{\rm e}=3$ keV, consistently with previous results \citep{farinelli08}.
The case of spherical geometry provides statistically equivalent results with $\tau_{\rm sph} \sim 2 \tau_{\rm slab}$.

\begin{figure}
\centering 
\includegraphics[angle=-90,width=1\columnwidth]{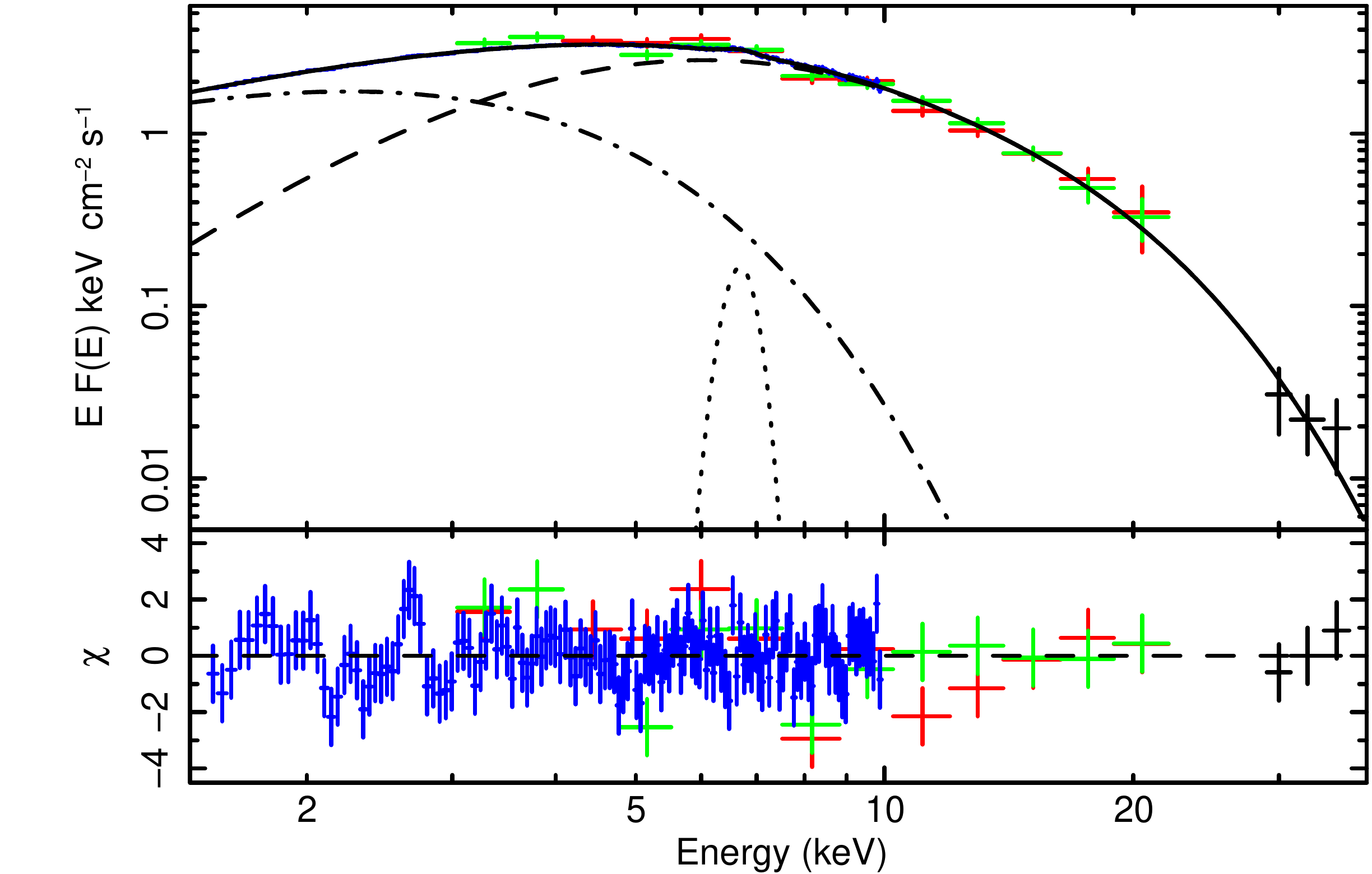}
\caption{\textit{Top panel:} Deconvolved absorption-corrected \nicer and \mbox{\integral}  spectrum of \cygx and the best-fitting model. The model parameters and errors are reported in Table~\ref{tab:model_fit}. 
Different colours of data points refer to \nicer\ ({blue}), \jmxone ({red}), \jmxtwo ({green}), and \isgri\ ({black}).
The model components are \diskbb\ ({dotted-dashed line}),
gaussian line ({dotted line}), and \comptt ({dashed line}). The \nicer points are not visible because their
error is smaller than the thickness of the total model line.
\textit{Bottom panel:} Residuals between the data and the model in units of $\sigma$.
For the feature around 2.6 keV, see  discussion in Sect. \ref{sect:spectral analysis}.  }
\label{eeuf_dbbcomptt}
\end{figure}

It is worth mentioning that unlike L22, who performed spectral analysis starting from 0.5 keV, we model the interstellar absorption with the simple \tbabs\ model in \xspec.
The residuals between the data and the model of the \nicer\ spectrum in the range 2--3 keV show some narrow features, resembling an absorption edge at 2 keV and an emission line around 2.7 keV. 
However, they are marginal and do not affect significantly the goodness of $\chi^2$ (see Fig.~\ref{eeuf_dbbcomptt}).
Indeed, we also tried a fit leaving free the relative abundances of oxygen and iron  (\tbfeo\ model in \xspec), but despite providing smaller residuals, no real improvement was observed in the  $\chi^2$ and the two additional parameters remained largely unconstrained. 

On the other hand, the \nicer\ data shows clear evidence of the emission iron line at $\sim 6.7$ keV, which was  
detected in  \bepposax\  and \nustar\ observations (see references in Sect.~\ref{sect:introduction}), and
that we modelled by adding to the continuum the \gauss\ component.

The best-fitting parameters are presented in Table~\ref{tab:model_fit}, while the model residuals and the deconvolved spectrum $EF_E$ are shown in Fig.~\ref{eeuf_dbbcomptt}.
The parameters of the Comptonized component are typical of the soft state of LMXBs, while the uncertainty in the value of the inner disc radius derived from \comptt, assuming an inclination angle of $i=60\degr$, is a factor of two, because the \diskbb\ model assumes the radial emissivity profile $T(R) \propto R^{-3/4}$, valid only in the outer part of accretion disc \citep{ss73, pt74}. 

We used  \ixpe\ data only for the polarimetric studies, and did not include the spectra in the  joint analysis  with \nicer\ and \integral.
The reason resides in a not yet proper correction of the telescope vignetting in the response matrix, leading in turn to a slightly different slope in the spectra of each DU. 
This effect was still present at the time of the (off-axis) pointing of \cygx, and becomes noticeable for a source with its brightness ($\sim 340$ mCrab in the 2--8 keV energy band). 
 
Nevertheless, skipping \ixpe\ for the spectral fit has no impact on the results because its energy band is fully covered by \nicer (see Fig.~\ref{eeuf_dbbcomptt}).
On the other hand, as the Stokes parameters $Q/I$ and $U/I$ are normalized quantities of photon counts, their derivation (as well as PD and PA) is not affected by different changes of efficiency over energy for the single DUs and  the results can be safely summed.
 
\begin{table}
\centering 
\caption{Best-fitting spectral parameters of the model \tbabs *(\diskbb+\comptt+\gauss) used to describe the joint \nicer and \integral\ spectrum of \cygx.
The errors are at the 90 per cent confidence level for a single parameter ($\Delta \chi^2 = 2.71$).}
\label{tab:model_fit}
\begin{tabular}{ll} 
 \hline
 \hline
  Parameter  & Value \\ 
 \hline  
 $N_{\rm H} (\times 10^{22} ~{\rm cm}^{-2}) $ & 0.12$^{+  0.04}_{-  0.04}$ \\ 
 $kT_{\rm in}$ (keV) &  0.95$^{+  0.23}_{-  0.11}$ \\ 
 
 $R_{\rm in}$ (km)\ $^a$  & 18$^{+5}_{-5}$  \\ 
 $kT_{\rm 0}$ (keV) & 1.23$^{+  0.22}_{-  0.10}$ \\ 
 $kT_{\rm e}$ (keV) & [3]  \\ 
 $\tau$\ $^b$ &  4.0$^{+  0.1}_{-  0.2}$ \\ 
 $N_{\rm comptt}$ &  0.40$^{+  0.15}_{-  0.17}$   \\ 
 \hline
 
 $E_{\rm ga}$ (keV) & 6.67$^{+  0.05}_{-  0.05}$  \\
 $\sigma_{\rm ga}$ (keV) & 0.29$^{+  0.08}_{-  0.06}$ \\
 $N_{\rm ga}$ $(\times 10^{-3})$      &  2.78$^{+1.01}_{-0.78}$  \\ 
 $EW$ (eV) & 40$^{+  10}_{-  10}$  \\
 \hline
 $\chi^2$/dof &  159/155\\
 $F_{\rm disc}^{\rm ph}/F_{\rm tot}^{\rm ph}$\ $^c$    & 0.44 \\
 $F_{\rm disc}^{\rm ene}/F_{\rm tot}^{\rm ene}$\ $^d$   & 0.47 \\
 $F_{\rm X}$\ $^e$   & $1.21\times 10^{-8}$ \\
 \hline
\end{tabular}
\begin{tablenotes}
\item $^a$ Assuming a distance of 7 kpc and an inclination angle of $i=60\degr$.  
\item $^b$ Obtained with slab geometry flag in \comptt. 
\item $^c$ Percentage of disk photon flux in the energy range 2--8 keV. 
\item $^d$ Percentage of disk energy flux in the energy range 0.1--50 keV.
\item $^e$ Model flux (ergs~cm$^{-2}$~s$^{-1}$) in the energy range 0.1--50 keV.
\end{tablenotes}
\end{table}

\subsection{Polarimetric analysis}
\label{polarimetric_analysis}

\begin{figure}
\includegraphics[angle=0,width=0.9\columnwidth]{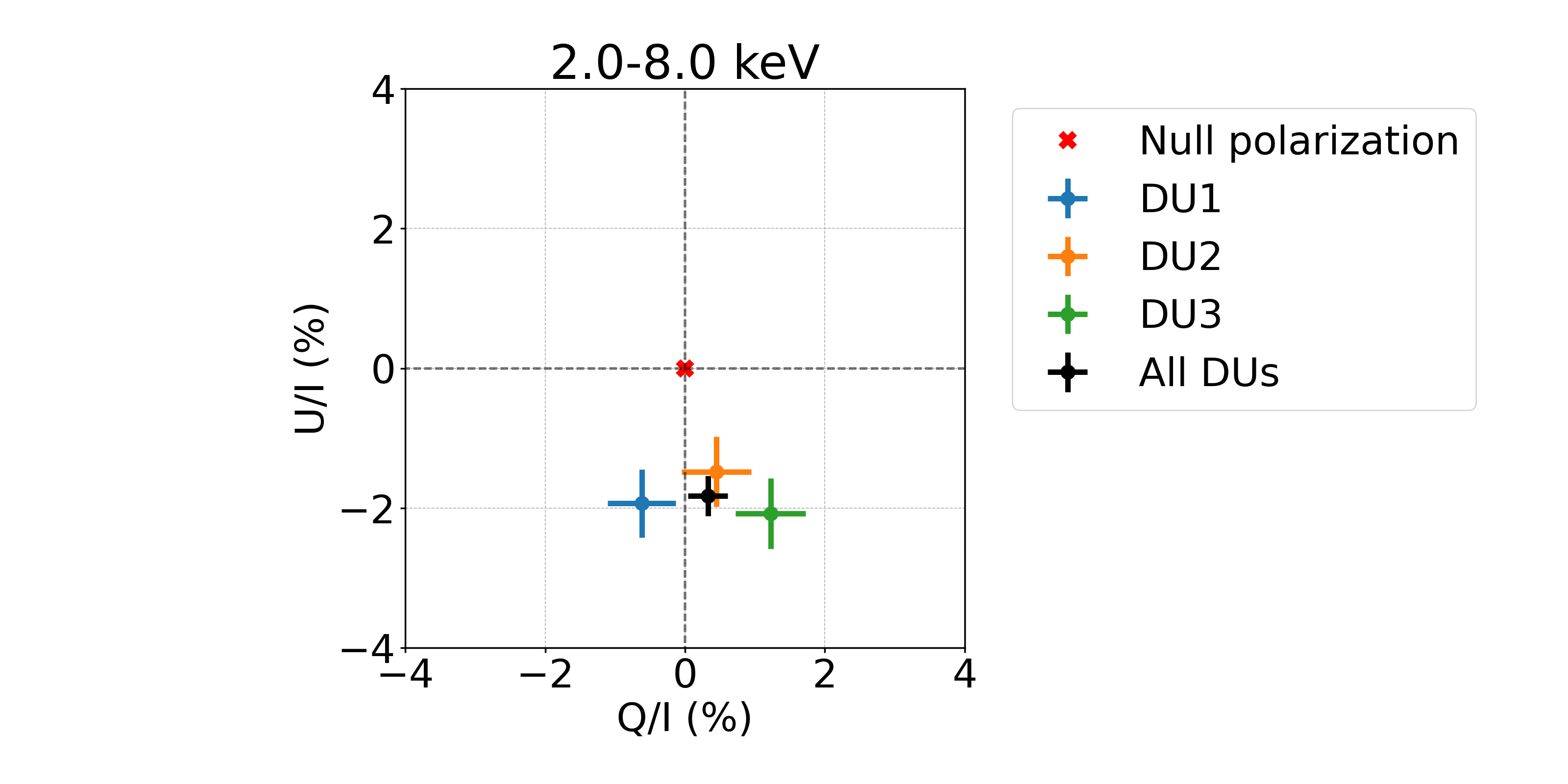}
\includegraphics[angle=0,width=0.8\columnwidth]{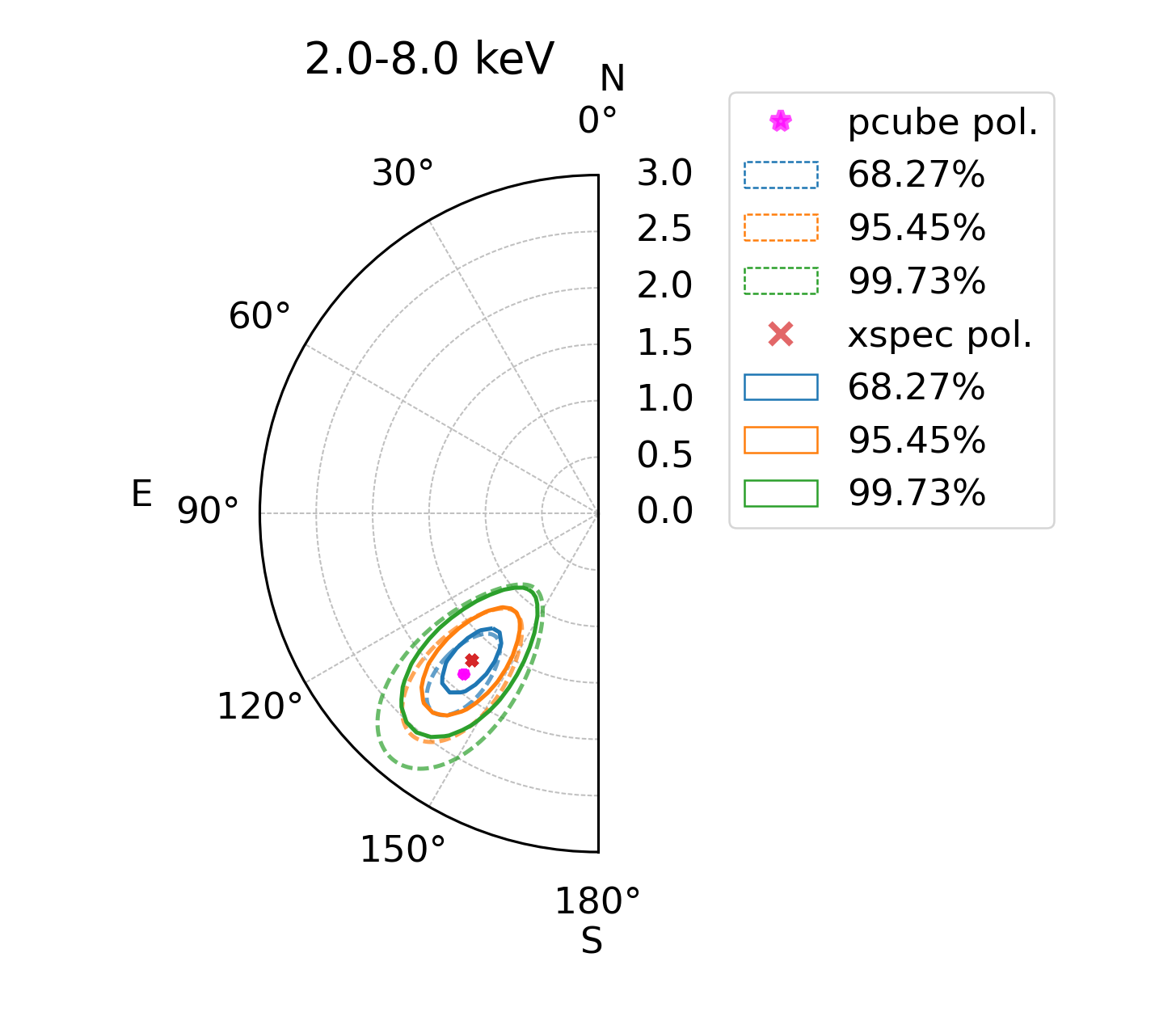}
\caption{\emph{Top panel}: Stokes parameters in the 2--8~keV energy band for each \ixpe\ DU and by summing all events. 
\emph{Bottom panel}: contour of PD and PA  at $68.27$,  $95.45$ and $99.73$ per cent  confidence levels obtained with \xspec\ (red cross and solid contours) and \pcube\ (pink star and dashed contours) by summing the events from the three DUs.}
\label{fig:p_ang_2-8kev}
\end{figure}

We have followed two different approaches for the analysis of the polarimetric properties of \cygx.
The first one is based on a model-independent analysis performed with the tool \pcube, with Stokes parameters for each DU and  total counts in different energy bands reported in Table~\ref{tab:stokes}. 
The same results  are also reported graphically in the top panel of Fig.~\ref{fig:p_ang_2-8kev}. 
The Stokes parameters from events detected by each DUs  reveal the presence of significant polarization at $6.4\sigma$ in the signal when data from all three DUs are taken into account.  

\begin{table} 
\centering 
\caption{Normalized Stokes parameters of Cyg X-2 computed with the \pcube\ algorithm. Errors at 1$\sigma$ level are reported.
\label{tab:stokes}}
\begin{tabular}{lrrrr} 
 \hline
 \hline
& DU1& DU2& DU3& All DUs \\ \hline
\multicolumn{5}{c}{2--8 keV}   \\ \hline
$Q/I$  ($\%$)                & $-0.62\pm0.49$ & $0.45\pm0.50$  & $1.22\pm0.50$ & $0.33\pm0.29$ \\
$U/I$ ($\%$)                        & $-1.90\pm0.49$ & $-1.48\pm0.50$ & $-2.08\pm0.50$ & $-1.82\pm0.29$ \\ \hline
\multicolumn{5}{c}{2--4 keV}   \\ \hline
$Q/I$ ($\%$)               & $0.03\pm0.48$ & $0.56\pm0.50$  & $1.06\pm0.50$ & $0.53\pm0.28$ \\
$U/I$ ($\%$)             & $-1.63\pm0.48$ & $-1.13\pm0.50$ & $-1.41\pm0.50$ & $-1.39\pm0.28$ \\ \hline
\multicolumn{5}{c}{4--8 keV}   \\ \hline
$Q/I$  ($\%$)                & $-2.06\pm0.95$ & $0.20\pm0.97$  & $1.58\pm0.97$ & $-0.13\pm0.56$ \\
$U/I$  ($\%$)                   & $-2.61\pm0.95$ & $-2.26\pm0.97$ & $-3.53\pm0.97$ & $-2.79\pm0.56$ \\ \hline
\end{tabular}
\end{table}

The second method is model-dependent and is based on fitting the Stokes parameter ($I$, $Q$, $U$) data  with the \xspec. 
We considered the model \polconst*\tbabs*(\diskbb+\comptt+\gauss) used for the broad-band spectral analysis with addition of the \polconst components which carries the additional polarimetric information.
This model assumes that the whole spectrum has a constant (i.e., energy-independent) PD and PA. 
In principle both \polconst\ and the spectral parameters should be left free, but data statistics and especially the \ixpe\ limited energy band do not allow to tightly constrain all the parameters simultaneously.

\begin{figure}
\includegraphics[angle=0,width=0.9\columnwidth]{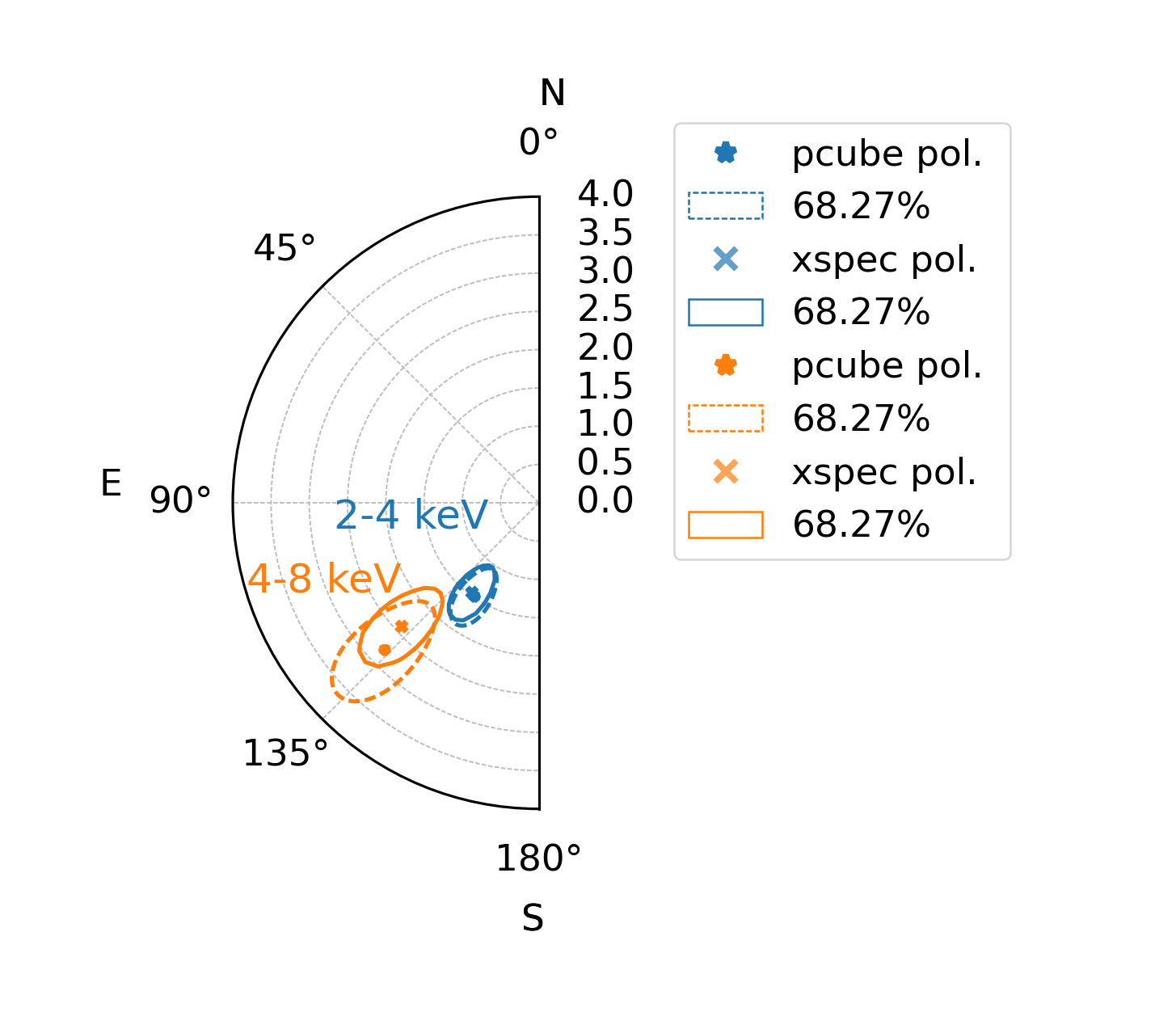}
\caption{Contour plot of PD and PA in the 2--4~keV (blue colour) and \mbox{4--8~keV} (orange colour) energy bands obtained with \xspec\ (cross central value and solid contours) and \pcube\ (star central value and dashed contours) by summing the events from the three DUs. 
Contours correspond to the $68.27$ per cent confidence level for the $\chi^2$ distribution with two degrees of freedom.}
\label{fig:p_ang_2-4-8kev}
\end{figure}

We thus have frozen all spectral parameters to the values obtained from the spectral fitting of the  \nicer+ \integral\ data (see Table~\ref{tab:model_fit}) leaving   only \polconst\ free.
The final results are given in Table~\ref{tab:polarization} in which the procedure described above is applied in the 2--8~keV, 2--4 keV and 4--8~keV energy bands together with \pcube\ results for a most direct comparison.
The same results are also graphically shown in Fig.~\ref{fig:p_ang_2-4-8kev}, and it is encouraging to observe the good agreement between the two methods, which strengthens the robustness of the measurements.

As expected, there is not a perfect overlap of the contour plots because of the methodological different approach. 
One caveat, for instance, may come from a simplified assumption of the energy-independent polarization in the \xspec\ model.
Whether this is true or not cannot be statistically inferred from the model-independent analysis with \ixpeobssim, as can be seen from the associated errors in Table~\ref{tab:polarization}.
Moreover, we outline again that the fit of the \ixpe\ spectra has been performed while keeping the continuum parameters frozen to the values obtained with \nicer\ and \integral. 
The vignetting issue of the telescope  might provide some slight (albeit not critical) differences in the derived spectral parameters which would be obtained from the \ixpe\ spectrum alone, in turn leading to a non-exact overlapping of the contour plots.


\begin{table} 
\centering
\caption{PD and PA of Cyg X-2 computed with both  \pcube\ and \xspec using the model \polconst*\tbabs*(\diskbb+\comptt+\gauss). 
Errors at 3$\sigma$ level for \pcube\  are estimated assuming Gaussian noise, whereas \xspec\ uncertainties are derived with the command \texttt{error} for one parameter of interest at a confidence level of $99.73$ per cent.   
\label{tab:polarization}}
\begin{tabular}{|l|c|c} 
 \hline \hline
 & \pcube   & \xspec  \\
\hline
\multicolumn{3}{c}{2--8 keV}   \\ \hline
PD ($\%$) & 1.85$\pm$0.87 &  1.72$\pm$0.71 \\
PA (deg)& 140$\pm$12 &  139$\pm$12  \\ \hline
\multicolumn{3}{c}{2--4 keV}   \\ \hline
PD ($\%$) & 1.49$\pm$0.84   & 1.46$ \pm 0.85 $ \\
PA (deg)& 144$\pm$15  &  143$\pm$18  \\ \hline
\multicolumn{3}{c}{4--8 keV}   \\ \hline
PD ($\%$) & 2.79$\pm$1.68   & 2.41$\pm$1.30  \\
PA (deg) & 134$\pm$18  &  132$\pm$17\\ \hline
\end{tabular}
\end{table}

As a further step, we exploited the capability of \xspec\ to disentangle the relative contribution to the polarization signal from the two spectral components (soft and hard) which are observed in the X-ray spectrum.
We considered the simplest case of a model with energy-independent PD and PA (\polconst), which lead a model in the form \tbabs*(\polconst*\diskbb+\polconst*\comptt + \gauss) in \xspec\ terminology. 
Again, we fixed the parameters of the spectral model to the best-fitting values reported in Table~\ref{tab:model_fit}.

\begin{figure}
\centering 
\includegraphics[angle=0,width=0.9\columnwidth]{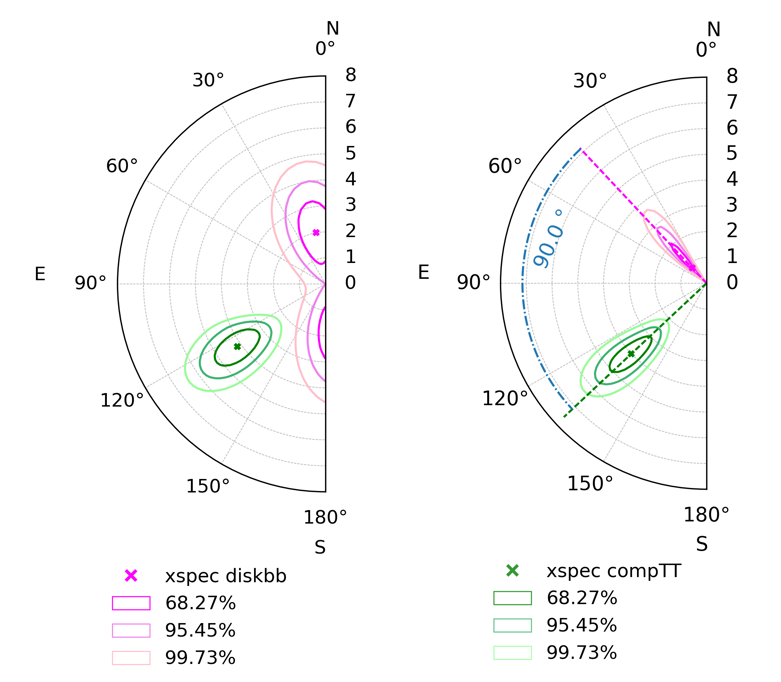}
\caption{Contour plot of PD and PA in the 2--8~keV energy band obtained with \xspec. 
The data have been fitted with two \polconst models separately for the \diskbb (pink colours) and \comptt (green colours) components. 
\textit{Left panel:} The PA  of \diskbb and \comptt are left free. 
\textit{Right panel:}  
The PA of \diskbb  was assumed to differ from the PA of \comptt by 90\degr. 
Contours correspond to the $68.27$, $95.45$ and $99.73$ per cent confidence levels.}
\label{fig:diskbb_compTT_angle}
\end{figure}

The contour plots are shown in Fig.~\ref{fig:diskbb_compTT_angle} (left panel), while the polarimetric parameters are reported in Table~\ref{tab:components_polarization}. 
We find  that the  PD and PA of the \comptt\ component are well constrained,  while the polarization of the \diskbb\ component is consistent with zero within the $95.45$ per cent  confidence level. 
However, there is a hint that the PA of  \diskbb\ is nearly perpendicular to that of the  \comptt, which is expected theoretically (see Sect.~\ref{sect.:discussion}).  
We thus allowed to vary only the PA of the \comptt component, with the PA of \diskbb\ being linked  by a relation $\textrm{PA}_{\textrm{\diskbb}}=\textrm{PA}_{\textrm{\comptt}}-90\degr$. 
The results with this specific setup are reported in Fig.~\ref{fig:diskbb_compTT_angle} (right panel) and again in Table~\ref{tab:components_polarization}, where we note that uncertainties in \comptt\ PD do not vary significantly, while the upper limit on \diskbb\ PD decreases from about 4.4 to 3.3 per cent.

\begin{table}
\centering 
\caption{Values of PD and PA associated to \diskbb\ and \comptt\  obtained
by fitting the \ixpe\ spectra with two 
\polconst\ components, with spectral parameters frozen to their best-fitting values obtained with \nicer\ and \integral\ (see Table~\ref{tab:model_fit}). The PA of the \diskbb\ and \comptt\ were assumed to be either free or linked so that $\textrm{PA}_{\textrm{\diskbb}}=\textrm{PA}_{\textrm{\comptt}}-90\degr$. Uncertainties are estimated with the \xspec\  \texttt{error} command at $99.73$ per cent  confidence level.}
\label{tab:components_polarization}
\begin{tabular}{|c|c|c|c|} 
 \hline
 \hline
 & & PD ($\%$) &PA (deg) \\ \hline
Free  &\diskbb             & $<4.4 $  & unconstrained \\
$\textrm{PA}_{\textrm{\diskbb}}$ and $\textrm{PA}_{\textrm{\comptt}}$ &\comptt             & $4.2_{-1.7}^{+2.0}$  & $126\pm 14$\\ \hline
&\diskbb             & $<3.3 $  & $43 \pm 11 $ \\
$\textrm{PA}_{\textrm{\diskbb}}=\textrm{PA}_{\textrm{\comptt}}-90\degr$ &\comptt & $4.0_{-1.7}^{+2.0}$  & $133 \pm 11 $ \\ \hline
\end{tabular}
\end{table}

\section{Discussion}\label{sect.:discussion}

In this work, we report the significant detection by \ixpe\ of X-ray polarization from a weakly magnetized NS LMXB \cygx.
After decades of discussions about the physics and geometry of the accretion in these system, polarimetric observations can now provide a fundamental contribution, allowing to remove at least partially the degeneracy of proposed spectral models which describe the broad-band spectral of LMXBs, particularly in the soft state.

The most intriguing result so far is the value of the PA, which is within errors consistent with the direction of the radio jet, episodically occurring when the source is in the HB of the Z-track.
This is in line with recent observations of \scox\ with \polarlight\ in the 3--8 keV energy band by \citet{long2022}, who came to the same conclusion: polarization is in the direction of the axis of symmetry of the system.
Moreover, the PA obtained from model-independent analysis (\pcube algorithm) is fully consistent within errors with that of the hard (Comptonized) part of the spectrum as obtained by a two-component fit of the X-ray continuum (see Sect. \ref{polarimetric_analysis}).

Using Fourier-frequency resolved spectroscopy, it has been claimed that in the soft state of NS LMXBs the hard component originates in the BL (or rather spreading layer, SL) covering the NS surface up to some latitude (see references in Sect. \ref{sect:introduction}).
To the first approximation, the SL is perpendicular to the disc plane, and the PA is expected to be rotated by $90\degr$ with respect to that of polarized radiation (if any) emerging from the electron-scattering dominated disc atmosphere \citep{chandrasekhar1960,Sobolev63}.
Because we measure a PA consistent with the jet position angle,  we exclude the possibility that the polarization signal directly comes from the accretion disc or a BL which is coplanar with the disc.
Note also that the PD=4.3$\pm$0.8 per cent  measured for \scox\ in the 4--8 keV range by \citet{long2022} is not compatible with the disc scattering atmosphere at the low inclination angle of the source \citep{cherepashchuk2021}.

Do all these results indicate that the disc is not polarized at all? Not necessarily.
The spectro-polarimetric fit provides only an upper limit of 3 per cent  to the PD value (and thus is principle consistent with zero).
The opacity in the inner disc is dominated by the electron scattering \citep{ss73, pt74}, and polarized radiation is expected to come from there, the PD depending, apart from system inclination, 
on the radial extension of  the electron scattering dominated region and
the optical depth  \citep{kallman2001}.
It is important to bear in mind that the classical Chandrasekhar's result is valid for
initial radiation at the bottom of a scattering medium with $\tau \gg 1$. 
As shown in \cite{st85}, the PD for $\tau \la 2$ can be higher than the corresponding value for semi-infinite atmosphere with the  polarization vector being aligned with the surface normal. 
The spectro-polarimetric analysis for \cygx\ (see Fig.~\ref{fig:diskbb_compTT_angle}), provides a hint in favour of a disk atmosphere with $\tau > 1$ with PA at right angle to both the observed one and to that associated to the Comptonized component.

Another complementary possibility to have a PA aligned to the system symmetry axis is
 through reflection from the disc.  This has been clearly shown in simulations of BH accretion disc in the soft state where disc self-irradiation with scattering at the top surface induces a $90\degr$ rotation of the PA \citep{Schnittman2009}. 
While self-irradiation is important only for a spin parameter $a$ close to unity, not achieved in NS LMXBs, the SL itself at the NS surface can serve as a source of photons. 
The reflection spectrum is produced essentially by photons scattered once and twice and these photons are highly polarized \citep{Matt93,Poutanen96} contributing significantly to the polarization signal even if the fraction of the reflection component to the total flux is small \citep{iaria2016, mondal2020}.
A geometry where the SL illuminates the accretion disk has been studied by \citet{Lapidus85}, who predicted a PD up to 6 per cent for an inclination angle of $i \approx 70\degr$ for  X-ray bursters during the phase
between bursts. This value is however overestimated as it does not consider the direct disk contribution,
which tends to produce a net lower polarization signal, unless the disk is polarized with PA  perpendicular
to the disk plane, as discussed above.

 The present spectroscopic data quality does not allow us to perform a thorough investigation of the contribution of the reflection, which instead has been studied  by \citet[][ hereafter M18]{mondal2018} and L22 using \nustar\ observations.
Their results show, however, that estimation of the source inclination as well as the disc inner radius from spectral analysis should be taken carefully because these parameters are strongly model-dependent.
Indeed, while M18 found $i \sim 20^{\circ}$ for both dipping and non dipping states occurring in the observation of 2015-01-07, L22 newly analysing the \nustar data of M18 with two additional observations performed on  2019-09-10 and 2019-09-12 joint with \nicer, reported a value of $i \sim 60^{\circ}$ for a non-dipping state,  more consistent with the optical data.
Further, M18 considered the model \diskbb plus \comptt (like in our fit) plus illumination of the accretion disc by a blackbody convolved with a relativistic disc kernel (\relconv * \reflionx), while 
 the parameters reported by L22 have been obtained  with two different components in which the illuminating source is either a blackbody (\relxillns) or a Comptonized blackbody (\rfxconv), but with the direct Comptonized spectrum described by a power law instead of \comptt.

We do not attempt to provide an estimation
of the system inclination based on the \ixpe\ polarimetric data, rather we point out that the value $P \ga 2$ per cent  (at $99.73$ per cent confidence level) as obtained from the \xspec\ modelling for the \comptt\ component  does not imply that it comes necessarily from the BL/SL at all. 
If polarization potentially comes from three regions (direct and reflected radiation from disc, plus BL) and spectro-polarimetric modelling is performed with only two components as in the present analysis, 
the situation is mathematically equivalent to having a system of two equations and three unknowns. 
The problem becomes degenerate and \xspec cannot do anything but attribute globally to the hard component (\comptt) a polarization which actually comes both directly from the BL/SL and from the fraction of radiation which is reflected from the disc before reaching the observer.
The marginal hint of increase of the PD with energy (see Fig.~\ref{fig:p_ang_2-4-8kev}) albeit to be taken cautiously, could be explained if the softer accretion disc is weaker polarized (and/or at right angle) than the harder BL/SL component. 

It is worth noticing that \cygx\ is a highly
variable source on time scales comparable or even
shorter than the duration of the \ixpe\ observation, as shown in Figs. \ref{maxi_lc} and \ref{jmx_lc} 
as well as in more recent \nustar\ analysis by M18 and L22.
The measured PD and PA reported in this work are averaged values over a period of about 2 days (albeit with 50\% of net exposure, see Table~\ref{tab:log_obs}) during which no flaring activity was detected and the source traced a NB-like pattern in the CCD. 
We also claim that the measured polarization parameters probably are not the average between, e.g., two extreme boundaries but they are rather stable for the source status caught during the \ixpe\ exposure.   
On the other hand, any dependence on the orbital phase should be less significant, 
because the \ixpe\ exposure covered  about 20 per cent of the source orbital period of $P_{\rm orb}=9.8$\,d.
Nevertheless, phase-dependent studies for  NS-LMXBs are expected to be of great interest, particularly to investigate deviations from the azimuthal symmetry induced by disc-warping effects \citep{abarr2020} or misalignment between the orbital and disc plane, as recently proposed by \citet{krawczynski2022} to explain \ixpe\ results of Cyg X-1. 
Our claims about disc reflection and SL contribution to the observed PD and PA need detailed Monte Carlo simulations which we are currently implementing and will be presented in a forthcoming paper.
The codes for computing  polarization of radiation have been focused mostly on accretion discs, where the central object is implicitly or explicitly assumed to be a BH \citep[e.g.][]{st85, ps96,schnittman2010, dovciak2011, schnittman2013, podgorny2022,Loktev22}, and for polar caps of accreting millisecond X-ray pulsars \citep{viironen2004,poutanen2020}.
To our knowledge, the first systematic attempt to investigate polarization properties in NS LMXBs using Monte Carlo simulations has been performed by \citet{gnarini2022}, who assumed however that the disc absorbs all the impinging radiation, and thus neglecting contribution to polarization from reflection by the disk photosphere.

The presence of such a gap in modelling  emission and polarization of LMXBs hosting a NS is probably motivated by the fact that while for BH systems Einstein's field equations have exact solutions (the Schwarzschild or Kerr metric) and efforts are focused only on the accretion disc properties,  for NS systems, even assuming simple Schwarzschild spacetime, the presence of a strong radiation pressure from the NS and the influence of the hard surface on the accretion disc structure makes it more difficult to construct a model of the accretion flow in the NS vicinity.
 
\section{Conclusions}

In this work, we have reported the first detection by the \ixpe\ satellite in the 2--8 keV energy band of X-ray polarization in the weakly magnetized NS LMXB \cygx.
The direction of the polarization angle is consistent with that of the source radio jet, which is presumably perpendicular to the accretion disc.
These measurements exclude the accretion disc and the BL coplanar with the disc as the main source of the polarized radiation while favouring the SL at the NS surface.
The data can also be consistent with reflection from the inner disc of the Comptonized radiation emerging from the SL itself. 
The data statistics, however, did not allow us to investigate the presence of a reflection component in the continuum, albeit we have indirect evidence for it from the presence of an iron fluorescence line.
The relative contribution of all these components needs Monte Carlo simulations which will be presented elsewhere.

The measured values of PD and PA allowed us to put strong (albeit not definitive) constraints on the accretion geometry of \cygx, and presumably by deduction to other NS LMXBs in the soft state.
Future \ixpe\ observations of other bright atoll- and Z-sources will be helpful in providing a more complete scenario of the accretion geometry of these sources in their soft state.

\section*{Acknowledgements}

The Imaging X-ray Polarimetry Explorer (\ixpe) is a joint US and Italian mission. The US contribution is supported by the National Aeronautics and Space Administration (NASA) and led and managed by its Marshall Space Flight Center (MSFC), with industry partner Ball Aerospace (contract NNM15AA18C). 

The Italian contribution is supported by the Italian Space Agency (Agenzia Spaziale Italiana, ASI) through contract ASI-OHBI-2017-12-I.0, agreements ASI-INAF-2017-12-H0 and ASI-INFN-2017.13-H0, and its Space Science Data Center (SSDC) with agreements ASI-INAF-2022-14-HH.0 and ASI-INFN 2021-43-HH.0, and by the Istituto Nazionale di Astrofisica (INAF) and the Istituto Nazionale di Fisica Nucleare (INFN) in Italy.

This research used data products provided by the \ixpe Team (MSFC, SSDC, INAF, and INFN) and distributed with additional software tools by the High-Energy Astrophysics Science Archive Research Center (HEASARC), at NASA Goddard Space Flight Center (GSFC). 
JP and SST were supported by the Russian Science Foundation grant 20-12-00364 and the Academy of Finland grants 333112, 349144, 349373, and 349906.

\section*{Data Availability}
The data used in this paper are publicly available in the HEASARC database.

\bibliographystyle{mnras}
\bibliography{cygx2_polarimetry}  

\clearpage 
\noindent \textbf{Affiliations:} \\
\noindent
$^{1}$INAF -- Osservatorio di Astrofisica e Scienza dello Spazio di Bologna, Via P. Gobetti 101, I-40129 Bologna, Italy\\
$^{2}$INAF -- IAPS, via del Fosso del Cavaliere 100, I-00113 Roma, Italy\\
$^{3}$Department of Physics and Astronomy, FI-20014 University of Turku, Finland\\
$^{4}$Space Research Institute of the Russian Academy of Sciences, Profsoyuznaya Str. 84/32, 117997 Moscow, Russia\\
$^{5}$Dipartimento di Matematica e Fisica, Universit\`a degli Studi Roma Tre, Via della Vasca Navale 84, I-00146 Roma, Italy\\
$^{6}$Department of Astronomy, University of Geneva, Ch. d'Ecogia 16, 1290, Versoix, Geneva, Switzerland\\
$^{7}$INAF -- Osservatorio Astronomico di Cagliari, via della Scienza 5, I-09047 Selargius (CA), Italy\\
$^8$University of New Hampshire, Department of Physics \& Astronomy and Space Science Center, 8 College Rd, Durham, NH 03824, USA\\
$^{9}$Instituto de Astrof\'isica de Andaluc\'ia-CSIC, Glorieta de la Astronom\'ia s/n, 18008, Granada, Spain\\ 
$^{10}$INAF Osservatorio Astronomico di Roma, Via Frascati 33, 00078 Monte Porzio Catone (RM), Italy\\ 
$^{11}$Space Science Data Center, Agenzia Spaziale Italiana, Via del Politecnico snc, 00133 Roma, Italy\\ 
$^{12}$Istituto Nazionale di Fisica Nucleare, Sezione di Pisa, Largo B. Pontecorvo 3, 56127 Pisa, Italy\\ 
$^{13}$Dipartimento di Fisica, Universit\`a di Pisa, Largo B. Pontecorvo 3, 56127 Pisa, Italy\\ 
$^{14}$NASA Marshall Space Flight Center, Huntsville, AL 35812, USA\\ 
$^{15}$Istituto Nazionale di Fisica Nucleare, Sezione di Torino, Via Pietro Giuria 1, 10125 Torino, Italy\\ 
$^{16}$Dipartimento di Fisica,  Universit\`a degli Studi di Torino, Via Pietro Giuria 1, 10125 Torino, Italy\\ 
$^{17}$INAF Osservatorio Astrofisico di Arcetri, Largo Enrico Fermi 5, 50125 Firenze, Italy\\ 
$^{18}$Dipartimento di Fisica e Astronomia,  Universit\`a degli Studi di Firenze, Via Sansone 1, 50019 Sesto Fiorentino (FI), Italy\\ 
$^{19}$Istituto Nazionale di Fisica Nucleare, Sezione di Firenze, Via Sansone 1, 50019 Sesto Fiorentino (FI), Italy\\ 
$^{20}$ASI - Agenzia Spaziale Italiana, Via del Politecnico snc, 00133 Roma, Italy\\ 
$^{21}$Istituto Nazionale di Fisica Nucleare, Sezione di Roma `Tor Vergata', Via della Ricerca Scientifica 1, 00133 Roma, Italy\\ 
$^{22}$Department of Physics and Kavli Institute for Particle Astrophysics and Cosmology, Stanford University, Stanford, California 94305, USA\\ 
$^{23}$Institut f\"ur Astronomie und Astrophysik, Universit\"at T\"ubingen, Sand 1, 72076 T\"ubingen, Germany\\ 
$^{24}$Astronomical Institute of the Czech Academy of Sciences, Bo\v{c}n\'{i} II 1401/1, 14100 Praha 4, Czech Republic\\ 
$^{25}$RIKEN Cluster for Pioneering Research, 2-1 Hirosawa, Wako, Saitama 351-0198, Japan\\ 
$^{26}$California Institute of Technology, Pasadena, CA 91125, USA\\ 
$^{27}$Yamagata University,1-4-12 Kojirakawa-machi, Yamagata-shi 990-8560, Japan\\ 
$^{28}$Osaka University, 1-1 Yamadaoka, Suita, Osaka 565-0871, Japan\\ 
$^{29}$University of British Columbia, Vancouver, BC V6T 1Z4, Canada\\ 
$^{30}$Department of Physics, Faculty of Science and Engineering, Chuo University, 1-13-27 Kasuga, Bunkyo-ku, Tokyo 112-8551, Japan\\ 
$^{31}$Institute for Astrophysical Research, Boston University, 725 Commonwealth Avenue, Boston, MA 02215, USA\\ 
$^{32}$Department of Astrophysics, St. Petersburg State University, Universitetsky pr. 28, Petrodvoretz, 198504 St. Petersburg, Russia\\ 
$^{33}$Physics Department and McDonnell Center for the Space Sciences, Washington University in St. Louis, St. Louis, MO 63130, USA\\ 
$^{34}$Finnish Centre for Astronomy with ESO,  20014 University of Turku, Finland\\ 
$^{35}$Universit\'{e} de Strasbourg, CNRS, Observatoire Astronomique de Strasbourg, UMR 7550, 67000 Strasbourg, France\\ 
$^{36}$MIT Kavli Institute for Astrophysics and Space Research, Massachusetts Institute of Technology, 77 Massachusetts Avenue, Cambridge, MA 02139, USA\\ 
$^{37}$Graduate School of Science, Division of Particle and Astrophysical Science, Nagoya University, Furo-cho, Chikusa-ku, Nagoya, Aichi 464-8602, Japan\\ 
$^{38}$Hiroshima Astrophysical Science Center, Hiroshima University, 1-3-1 Kagamiyama, Higashi-Hiroshima, Hiroshima 739-8526, Japan\\ 
$^{39}$Department of Physics, The University of Hong Kong, Pokfulam, Hong Kong\\ 
$^{40}$Department of Astronomy and Astrophysics, Pennsylvania State University, University Park, PA 16802, USA\\ 
$^{41}$Universit\'{e} Grenoble Alpes, CNRS, IPAG, 38000 Grenoble, France\\ 
$^{42}$Center for Astrophysics, Harvard \& Smithsonian, 60 Garden St, Cambridge, MA 02138, USA\\ 
$^{43}$INAF Osservatorio Astronomico di Brera, Via E. Bianchi 46, 23807 Merate (LC), Italy\\ 
$^{44}$Dipartimento di Fisica e Astronomia, Universit\`a degli Studi di Padova, Via Marzolo 8, 35131 Padova, Italy\\ 
$^{45}$Dipartimento di Fisica, Universit\`a degli Studi di Roma `Tor Vergata', Via della Ricerca Scientifica 1, 00133 Roma, Italy\\ 
$^{46}$Department of Astronomy, University of Maryland, College Park, Maryland 20742, USA\\ 
$^{47}$Mullard Space Science Laboratory, University College London, Holmbury St Mary, Dorking, Surrey RH5 6NT, UK\\ 
$^{48}$Anton Pannekoek Institute for Astronomy \& GRAPPA, University of Amsterdam, Science Park 904, 1098 XH Amsterdam, The Netherlands\\ 
$^{49}$Guangxi Key Laboratory for Relativistic Astrophysics, School of Physical Science and Technology, Guangxi University, Nanning 530004, China\\

\bsp	
\label{lastpage}

\end{document}